\documentclass[%
 reprint,
 amsmath,amssymb,
 aps,
]{revtex4-2}

\usepackage{graphicx}
\usepackage{dcolumn}
\usepackage{bm}

\usepackage{amsmath}
\usepackage{amsthm}
\usepackage{braket}

\begin{document}

\preprint{APS/123-QED}

\title{Quantum tomography for arbitrary single-photon polarization-path states}

\author{J. L. Montenegro Ferreira}
\email{lukas.montenegrof@gmail.com}
 
\author{B. de Lima Bernardo}%
 \email{bertulio.fisica@gmail.com}
\affiliation{%
Departamento de Física, Universidade Federal da Paraíba, 58051-900 João Pessoa, PB, Brazil
}%

\date{\today}

\begin{abstract}
Quantum state tomography (QST), the process through which the density matrix of a quantum system is characterized from measurements of specific observables, is a fundamental pillar in the fields of quantum information and computation. In this work, we propose a simple QST method to reconstruct the density matrix of two qubits encoded in the polarization and path degrees of freedom of a single photon, which can be realized with a single linear-optical setup. We demonstrate that the density matrix can be fully described in terms of the Stokes parameters related to the two possibles paths of the photon, together with a quantum version of the two-point Stokes parameters introduced here. Our findings put forward photonic circuits for the investigation of the dynamics of open quantum systems.   
\end{abstract}

\maketitle


\section{\label{sec:level1}Introduction}

With the advent of quantum information theory we learned that entanglement is an essential resource for many important tasks, such as quantum communication \cite{bennet1}, quantum computation \cite{rauss}, quantum cryptography \cite{ekert}, and teleportation \cite{bennet2}. This fact has led to an intense search for ways to create and characterize entanglement in many different physical architectures, among which photons occupy a special place due to the capacity of carrying information through long distances \cite{couteau}. Notably, different degrees of freedom (DOF) can be used to entangle photons, including position, linear momentum, polarization, orbital angular momentum, frequency and time-bin \cite{kok,mair,marc}. Nevertheless, entanglement can also be created between the different DOF of a single photon, which provides an alternative way to achieve high-dimensional entangled states over the use of multiple photons entangled in a single DOF \cite{pilnyak}.        

\hfill

In order to benefit from the valuable resources extractable from quantum states, besides the physical implementation, we also need to be able to control, measure and characterize them. The process of reconstruction of a quantum state from measurements made on an ensemble of identical systems is called quantum state tomography (QST) \cite{ariano,lvov,toninelli}. The idea is to process the outcomes obtained from a  complete set of observables to identify all elements of the density matrix of the system. To characterize a $N$-qubit state, measurements of 2$^{2N}$ different observables are required \cite{james}. Evidently, this exponential relation sets a limit on the feasibility of experimental realizations of QST with correlated many-body systems. Although this fact, an alternative platform, which has not been much explored, is the application of QST to high-dimension states encoded in different DOF of the same particle. Some exceptions are the studies of single photons entangled in polarization and time and single neutrons entangled in spin and path  \cite{pilnyak,hasegawa}.  

In this work, we propose an experimental setup to realize QST for an arbitrary polarization-path state of a single photon, which encodes two qubits of information. The scheme is fundamentally based on the composition of a two-path interferometer and a set of four Stokes parameters measurers conveniently distributed. The tomographic protocol demands only a single arrangement of linear optical devices, in the sense that one does not need to modifify the experimental setup in order to measure all necessary observables. In what concerns our theoretical approach, besides the well known one-point Stokes parameters, we introduce a quantum-mechanical version of the two-point Stokes parameters, a concept so far only established for classical optical fields \cite{koro}. In fact, the polarization-path density matrix is fully reconstructed with basis on the quantum counterparts of the one- and two-point Stokes parameters, which we call one-path Stokes parameters (OPSP) and two-path Stokes parameters (TPSP), respectively. 

The paper is organized as follows. First, we outline the quantum description of the OPSP and propose a similar treatment for the TPSP. Then, we use the results to demonstrate a QST method to reconstruct an arbitrary polarization-path state of a single photon. Finally, we present our conclusions and discuss how our QST method could be employed as a new tool in the study of the dynamics of open single- and two-qubit systems with quantum optical experiments.  

\section{One- and two-point Stokes parameters: a quantum-mechanical viewpoint}

Suppose that photons from an ensemble propagate one at a time through two possible paths 0 and 1 whose states are defined as $\ket{0}$ and $\ket{1}$, and the polarization is described in terms of the horizontal and vertical states, $\ket{H}$ and $\ket{V}$. In this case, the polarization and path properties of the photons can be jointly described in terms of the polarization-path density matrix \cite{bert}, 
\begin{equation}\label{density}
\hat{\rho}=
\begin{pmatrix}
\rho_{11} & \rho_{12} & \rho_{13} & \rho_{14}\\
\rho_{21} & \rho_{22} & \rho_{23} & \rho_{24}\\
\rho_{31} & \rho_{32} & \rho_{33} & \rho_{34}\\
\rho_{41} & \rho_{42} & \rho_{43} & \rho_{44},
\end{pmatrix},
\end{equation}
which is written in the basis $\{\ket{H,0}; \ket{H,1}; \ket{V,0}; \ket{V,1}\}$. In this framework, we can use the OPSP to characterize the polarization of the photons in each path. From a quantum-mechanical viewpoint, these parameters are given by the ensemble average of the Pauli operators \cite{james,bert,santos}. As such, the OPSP for the photons that propagate through path 0 are given by    

\begin{subequations}
    \begin{equation}\label{s0p0}
s^{(0)}_{0}=\mathrm{Tr}[(\ket{H,0}\bra{H,0}+\ket{V,0}\bra{V,0})\hat{\rho}] = \rho_{11} + \rho_{33},
    \end{equation}
    \begin{equation}\label{s0p1}
        s^{(0)}_{1}=\mathrm{Tr}[(\ket{H,0}\bra{H,0}-\ket{V,0}\bra{V,0})\hat{\rho}] = \rho_{11} - \rho_{33},
    \end{equation}
    \begin{equation}\label{s0p2}
        s^{(0)}_{2}=\mathrm{Tr}[(\ket{H,0}\bra{V,0}+\ket{V,0}\bra{H,0})\hat{\rho}] = \rho_{13} + \rho_{31},
    \end{equation}
    \begin{equation}\label{s0p3}
        s^{(0)}_{3}=i\{\mathrm{Tr}[(\ket{V,0}\bra{H,0}-\ket{H,0}\bra{V,0})\hat{\rho}]\} = i(\rho_{13} - \rho_{31}),
    \end{equation}
\end{subequations}

where $\mathrm{Tr}[.]$ denotes the trace operation. Similarly, the polarization of the photons that propagate through path 1 is described by the corresponding OPSP:

\begin{subequations}
    \begin{equation}\label{s1p0}
s^{(1)}_{0}=\mathrm{Tr}[(\ket{H,1}\bra{H,1}+\ket{V,1}\bra{V,1})\hat{\rho}] = \rho_{22} + \rho_{44},
    \end{equation}
    \begin{equation}\label{s1p1}
        s^{(1)}_{1}=\mathrm{Tr}[(\ket{H,1}\bra{H,1}-\ket{V,1}\bra{V,1})\hat{\rho}] = \rho_{22} - \rho_{44},
    \end{equation}
    \begin{equation}\label{s1p2}
        s^{(1)}_{2}=\mathrm{Tr}[(\ket{H,1}\bra{V,1}+\ket{V,1}\bra{H,1})\hat{\rho}] = \rho_{24} + \rho_{42},
    \end{equation}
    \begin{equation}\label{s1p3}
        s^{(1)}_{3}=i\{\mathrm{Tr}[(\ket{V,1}\bra{H,1}-\ket{H,1}\bra{V,1})\hat{\rho}]\} = i(\rho_{24} - \rho_{42}).
    \end{equation}
\end{subequations}
From Eqs.~(\ref{s0p0}) to~(\ref{s1p3}) we observe that eight out of the sixteen entries of the polarization-path density matrix of Eq.~(\ref{density}) are sufficient to fully characterize the polarization of the photons propagating in each path.

\hfill

\begin{figure}
\begin{center}
    \includegraphics[width=0.49\textwidth]{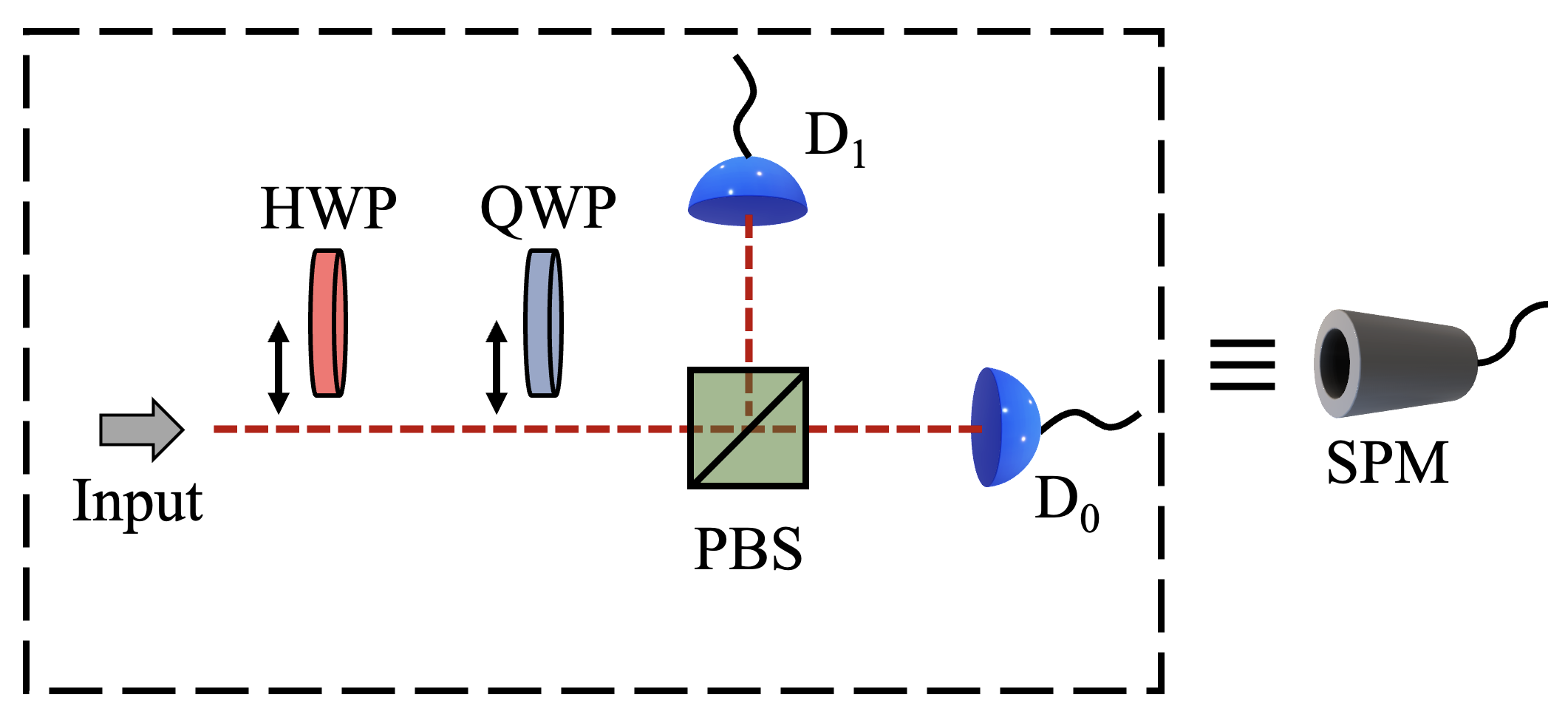}
     \caption{Experimental setup to measure the OPSP. Photons with arbitrary polarization propagate toward the polarizing beam splitter PBS. The parameter $s_{0}$ is found through the sum of the signals of the photodetectors D$_0$ and D$_1$. The other parameters are assessed through the difference of the signals: $s_{1}$ without the wave plates, $s_{2}$ only with the half-wave plate HWP, and $s_{2}$ only with the quarter-wave plate QWP. This apparatus is identified as a Stokes parameters measurer (SPM).}
    \label{OPSPmeasurer}
\end{center}
\end{figure}

The experimental setup to measure the OPSP is shown in Fig.~\ref{OPSPmeasurer} (see also Refs.~\cite{altepeter,bayra}). It consists of a polarizing beam-splitter PBS, which transmits $\ket{H}$ and reflects $\ket{V}$ states, a half-wave plate HWP, a quarter-wave plate QWP, and two photodetectors D$_0$ and D$_1$. The parameter $s_0$ is simply given by the sum of the number of photons registered in the photodetectors, independent of whether the wave plates are introduced along the optical path or not, divided by the number of input photons, $s_{0} = (N_0 + N_1)/ N_{in}$. The parameters $s_1$, $s_2$ and $s_3$ are given by the difference between the signals registered by D$_0$ and D$_1$ divided by the number of input photons, $s_{k} = (N_0 - N_1)/ N_{in}$ with $k=1,2,3$. Nevertheless, $s_1$ is measured without the wave plates, $s_2$ when only the HWP is inserted with fast axis at angle $\pi/8$ with respect to the horizontal, and $s_3$ when only the QWP is inserted with fast axis at angle $\pi/4$ with respect to the horizontal. The 2$\times$2 transformation matrices of the QWP and HWP are given by \cite{james}

\begin{equation}\label{QWP}
\hat{U}_{QWP}(\theta)= \frac{1}{\sqrt{2}}
\begin{pmatrix}
i + cos(2 \theta) & sin (2 \theta) \\
sin (2 \theta) & i - cos(2 \theta)
\end{pmatrix},
\end{equation}

\begin{equation}\label{HWP}
\hat{U}_{HWP}(\theta)= 
\begin{pmatrix}
cos(2 \theta) & sin (2 \theta) \\
sin (2 \theta) & - cos(2 \theta)
\end{pmatrix},
\end{equation}

where $\theta$ is the angle between the fast axis and the horizontal. The effect of $\hat{U}_{HWP}(\pi/8)$ is to convert $\ket{D}$ ($\ket{A}$) to $\ket{H}$ ($\ket{V}$), whereas for $\hat{U}_{QWP}(\pi/4)$ the effect is to convert $\ket{R}$ ($\ket{L}$) to $\ket{H}$ ($\ket{V}$). Here, we define $\ket{D} = 1/ \sqrt{2} (\ket{H} + \ket{V})$, $\ket{A} = 1/ \sqrt{2} (\ket{H} - \ket{V})$, $\ket{R} = 1/ \sqrt{2} (\ket{H} + i \ket{V})$ and $\ket{L} = 1/ \sqrt{2} (\ket{H} - i \ket{V})$. 

\hfill

Similar to the derivation of the OPSP in Eqs.~(\ref{s0p0}) to~(\ref{s1p3}), and based on the mathematical structure of the classical two-point Stokes parameters introduced in Ref.~\cite{koro}, the quantum generalization of these parameters can be defined under the present framework as  

\begin{subequations}
    \begin{equation}\label{Sp0}
        S_{0}=\mathrm{Tr}[(\ket{H,0}\bra{H,1}+\ket{V,0}\bra{V,1})\hat{\rho}]= \rho_{21}+\rho_{43},
    \end{equation}
    \begin{equation}\label{Sp1}
        S_{1}=\mathrm{Tr}[(\ket{H,0}\bra{H,1}-\ket{V,0}\bra{V,1})\hat{\rho}]= \rho_{21}-\rho_{43},
    \end{equation}
    \begin{equation}\label{Sp2}
        S_{2}=\mathrm{Tr}[(\ket{H,0}\bra{V,1}+\ket{V,0}\bra{H,1})\hat{\rho}]= \rho_{23}+\rho_{41},
    \end{equation}
    \begin{equation}\label{Sp3}
        S_{3}=i\{\mathrm{Tr}[(\ket{V,0}\bra{H,1}-\ket{H,0}\bra{V,1})\hat{\rho}]\}= i(\rho_{23}-\rho_{41}),
    \end{equation}
\end{subequations}

\hfill

which we call two-path Stokes paramters (TPSP). While the OPSP are real numbers, as a consequence of the Hermiticity of the polarization-path density matrix, the TPSP are generally complex and are obtained from correlations that involve photons propagating through both paths. In what follows we shall see how the TPSP can be measured, and their usefulness in realizing polarization-path state tomography.  

\section{Polarization-path State Tomography}

The results of Eqs.~(\ref{s0p0}) to~(\ref{s1p3}), together with Eqs.~(\ref{Sp0}) to~(\ref{Sp3}), allow us to rewrite the polarization-path matrix of Eq.~(\ref{density}) completely in terms of the OPSP and TPSP,  

\begin{equation}\label{density2}
\hat{\rho}= \frac{1}{2}
\begin{pmatrix}
s^{(0)}_0 + s^{(0)}_1 & S^{*}_0 + S^{*}_1 & s^{(0)}_2 - i s^{(0)}_3 & S^{*}_2 - iS^{*}_3\\
S_0 + S_1 & s^{(1)}_0 + s^{(1)}_1 & S_2 - i S_3 & s^{(1)}_2 - i s^{(1)}_3\\
s^{(0)}_2 + i s^{(0)}_3 & S^{*}_2 + i S^{*}_3 & s^{(0)}_0 - s^{(0)}_1 & S^{*}_0 - S^{*}_1\\
S_2 + i S_3 & s^{(1)}_2 + i s^{(1)}_3 & S_0 - S_1 & s^{(1)}_0 - s^{(1)}_1
\end{pmatrix}.
\end{equation}

\hfill

This equation tells us that, given an ensemble of photons described by an unknown polarization-path density matrix, if we are able to measure the OPSP and TPSP we can fully reconstruct the state. Therefore, the knowledge about these parameters allows us to realize polarization-path QST. To this end, we are left with the task of designing a way to assess the TPSP. 

\hfill

\begin{figure}
\begin{center}
    \includegraphics[width=0.49\textwidth]{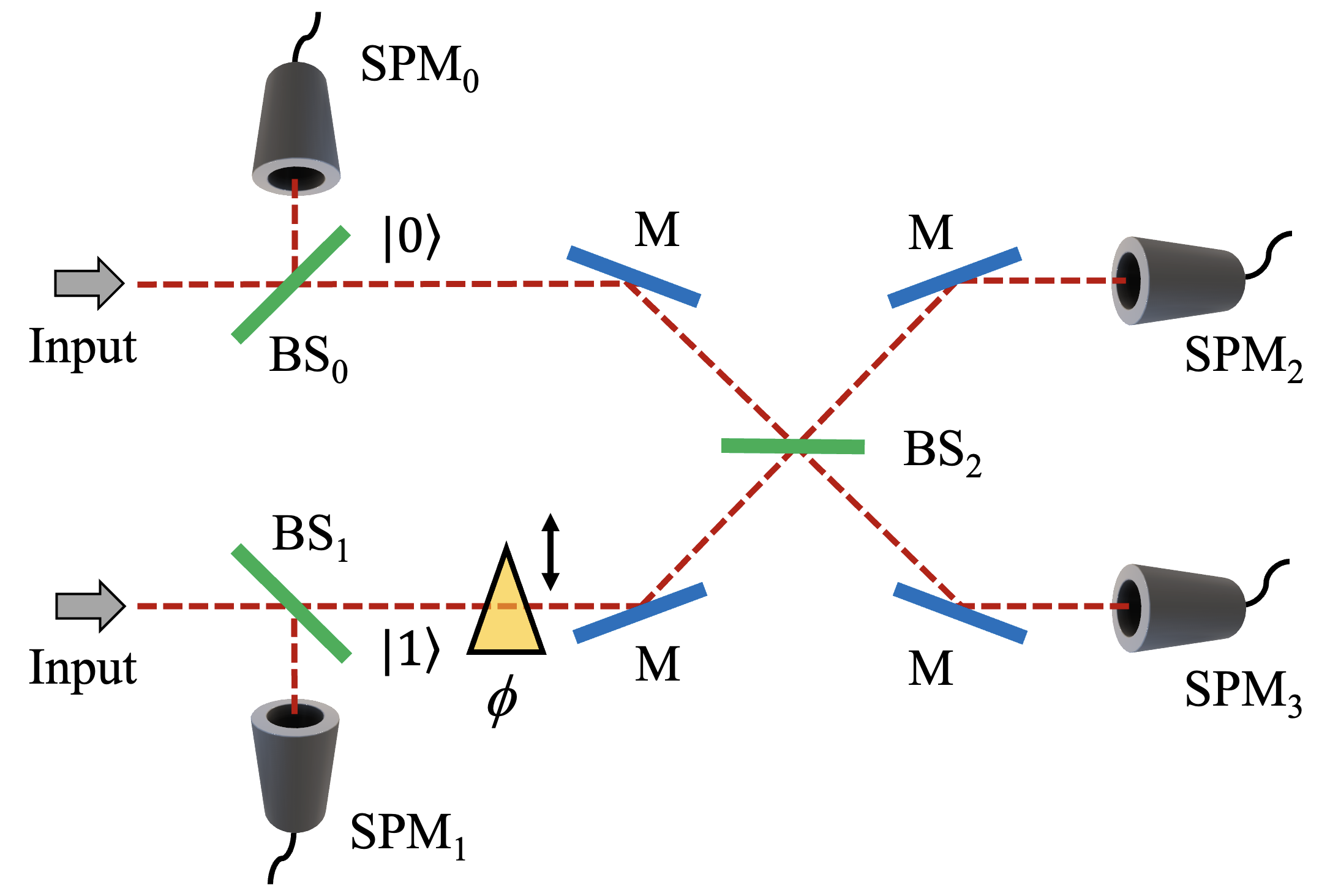}
     \caption{Experimental setup to realize a complete polarization-path state tomography. Photons enter the setup in a arbitrary state of polarization and path. The 1:1 beam splitters BS$_0$ and BS$_1$ reflect part of the photons toward the SPM$_0$ and SPM$_1$ apparatuses to perform OPSP measurements, as depicted in Fig.~\ref{OPSPmeasurer}. For the transmitted photons, an adjustable phase shift $\phi$ is applied in the lower path, after which the two paths are set as the input ports of the 1:1 beam splitter BS$_2$. Photons from the output ports are then collected by SPM$_2$ and SPM$_3$, which are used to obtain the TPSP. The four M represent perfectly reflecting mirrors.}
    \label{TPSPmeasurer}
\end{center}
\end{figure}

In order to interpret the physical meaning of the four TPSP we first consider the experimental setup sketched in Fig.~\ref{TPSPmeasurer}. It consists of single-photons with arbitrary polarization entering the apparatus through the input paths $\ket{0}$ and $\ket{1}$, so that we can consider the general input state as described by the polarization-path density matrix as in Eq.~(\ref{density}). Initially, one half of the photons are reflected by two lossless 1:1 beam splitters, BS$_{0}$ and BS$_{1}$, on which measurements of the OPSP can be made with SPM$_{0}$ and SPM$_{1}$, according to the scheme shown in Fig.~\ref{OPSPmeasurer}. Since one half of all input photons are used to this end, in this case the OPSP are identified according the relations $s_{0} = 2(N_0 + N_1)/ N_{in}$ and $s_{k} = 2(N_0 - N_1)/ N_{in}$ with $k=1,2,3$. Here, $N_{in}$ represents the total number of input photons in the experimental setup.

\hfill

For the photons transmitted by BS$_{0}$ and BS$_{1}$, we have that those propagating along path 1 are subjected a controllable phase shift $\phi$. Next, photons from both paths are superposed by a 1:1 lossless beam splitter BS$_{2}$, after which they follow toward SPM$_{2}$ and SPM$_{3}$. The effect of the phase shift is described by the operator $\hat{A} = \ket{0}\bra{0} + e^{i \phi} \ket{1}\bra{1}$ and the beam splitter by $\hat{B} = 1 / \sqrt{2} \left(\ket{0}\bra{0} - \ket{0}\bra{1} + \ket{1}\bra{0} + \ket{1}\bra{1} \right)$ \cite{nielsen}. Together they cause the unitary transformation $\hat{U} = \hat{I}_{p} \otimes \hat{B} \hat{A}$, where $\hat{I}_{p} = \ket{H}\bra{H} + \ket{V}\bra{V}$ is the identity operator in the polarization Hilbert space. 
The corresponding transformation matrix is then given as

\begin{equation}\label{transformation}
\hat{U}= \frac{1}{\sqrt{2}}
\begin{pmatrix}
1 & -e^{i \phi} & 0 & 0\\
1 & e^{i \phi} & 0 & 0\\
0 & 0 & 1 & -e^{i \phi}\\
0 & 0 & 1 & e^{i \phi}
\end{pmatrix}.
\end{equation}

After BS$_2$ the input state $\hat{\rho}$ is transformed into the final state $\hat{\rho}_{f} = \hat{U} \hat{\rho} \hat{U}^{\dagger}$. The calculation of $\hat{\rho}_{f}$ is straightforward, the result is another polarization-path state represented in the basis $\{\ket{H,0}; \ket{H,1}; \ket{V,0}; \ket{V,1}\}$, but it is too lengthy to be written out explicitly. For $\hat{\rho}_{f}$ we have for example that the basis state $\ket{H,0}$ ($\ket{V,1}$) represents a horizontally- (vertically-)polarized photon in the optical mode which is collected by SPM$_{3}$ (SPM$_{2}$), and so on.

\hfill

Next, with the procedure described in Fig.~\ref{OPSPmeasurer}, the OPSP for the state $\hat{\rho}_{f}$ in the paths $\ket{0}$ and $\ket{1}$ can be found. Given that one half of the input photons are transmitted by BS$_{0}$ and BS$_{1}$, these OPSP must also be identified according to the relations $s_{0} = 2(N_0 + N_1)/ N_{in}$ and $s_{k} = 2(N_0 - N_1)/ N_{in}$ with $k=1,2,3$. After some calculations, we can find that the OPSP for the photons in path $\ket{0}$ (i.e., those collected by SPM$_3$) varies as a function of $\phi$  according to the relations:

\begin{subequations}
    \begin{equation}\label{s0d0}
        s^{(0)}_{0f}(\phi) = \frac{1}{2}[s^{(0)}_{0} + s^{(1)}_{0} - 2\mathrm{Re}(S_{0}e^{i\phi})],
    \end{equation}
    \begin{equation}\label{s1d0}
        s^{(0)}_{1f}(\phi) = \frac{1}{2}[s^{(0)}_{1} + s^{(1)}_{1} - 2\mathrm{Re}(S_{1}e^{i\phi})],
    \end{equation}
    \begin{equation}\label{s2d0}
        s^{(0)}_{2f}(\phi) = \frac{1}{2}[s^{(0)}_{2} + s^{(1)}_{2} - 2\mathrm{Re}(S_{2}e^{i\phi})],
    \end{equation}
    \begin{equation}\label{s3d0}
        s^{(0)}_{3f}(\phi) = \frac{1}{2}[s^{(0)}_{3} + s^{(1)}_{3} - 2\mathrm{Re}(S_{3}e^{i\phi})],
    \end{equation}
\end{subequations}

\hfill

where $\mathrm{Re}(.)$ denotes the real part. Similarly, we can write the OPSP as a function of $\phi$ for the photons in path $\ket{1}$ (i.e., those collected by SPM$_2$) as follows: 

\begin{subequations}
    \begin{equation}\label{s0d1}
        s^{(1)}_{0f}(\phi) = \frac{1}{2}[s^{(0)}_{0} + s^{(1)}_{0} + 2\mathrm{Re}(S_{0}e^{i\phi})],
    \end{equation}
    \begin{equation}\label{s1d1}
        s^{(1)}_{1f}(\phi) = \frac{1}{2}[s^{(0)}_{1} + s^{(1)}_{1} + 2\mathrm{Re}(S_{1}e^{i\phi})],
    \end{equation}
    \begin{equation}\label{s2d1}
        s^{(1)}_{2f}(\phi) = \frac{1}{2}[s^{(0)}_{2} + s^{(1)}_{2} + 2\mathrm{Re}(S_{2}e^{i\phi})],
    \end{equation}
    \begin{equation}\label{s3d1}
        s^{(1)}_{3f}(\phi) = \frac{1}{2}[s^{(0)}_{3} + s^{(1)}_{3} + 2\mathrm{Re}(S_{3}e^{i\phi})].
    \end{equation}
\end{subequations}

We observe that, due to the unitarity of the interferometer transformation, the OPSP obey a complementary relation:
\begin{equation}
    s^{(0)}_{n} + s^{(1)}_{n} = s^{(0)}_{nf} + s^{(1)}_{nf},
\end{equation}

\hfill

with $n=0,1,2,3$. This relation, which holds for all values of $\phi$, reflects the fact that there are no losses in the interferometer, and the degree of polarization of the photons is conserved.

\hfill

At this point, we call attention to an important result obtained from Eqs.~(\ref{s0d0}) to~(\ref{s3d1}). The TPSP quantify the contrast in the interference behavior of the OPSP for the state $\hat{\rho}_f$ when $\phi$ is varied. A similar interpretation was also obtained in the classical context of Young's double-slit experiment \cite{setala}. This shows that by measuring the interference patterns of the OPSP with the interferometer of Fig.~\ref{TPSPmeasurer}, we can assess the TPSP in a simple form. As we shall see, this will be useful to complete the characterization of the state $\hat{\rho}$. Let us first demonstrate in more detail how to obtain the TPSP. We have that $\mathrm{Re}(S_n e^{i \phi}) = \mathrm{Re}(S_n) \cos \phi - \mathrm{Im}(S_n) \sin \phi$, where $\mathrm{Im}(.)$ denotes the imaginary part, with $n=0,1,2,3$. Therefore, to obtain the real and imaginary parts of the TPSP, we only need to obtain the OPSP along one of the path states of $\hat{\rho}_f$ for $\phi = 0$ and $\phi = \pi /2$. In considering path $\ket{0}$, from Eqs.~(\ref{s0d0}) to~(\ref{s3d0})
we can find that:
\begin{equation}
\label{real0}
\mathrm{Re}(S_n) = \frac{1}{2}[s^{(0)}_n + s^{(1)}_n] - s^{(0)}_{nf}(0),    
\end{equation}
\begin{equation}
\label{imaginary0}
\mathrm{Im}(S_n) = - \frac{1}{2}[s^{(0)}_n + s^{(1)}_n] + s^{(0)}_{nf}(\pi/2).  
\end{equation}
From these two relations we can directly determine the TPSP, $S_n = \mathrm{Re}(S_n) + i \mathrm{Im}(S_n)$, which means that they can be determined through simple OPSP measurements. Additionally, in order to realize the state tomography with the least number of input photons, it is also important to get information about the TPSP with the photons that propagate along path $\ket{1}$. From Eqs.~(\ref{s0d1}) to~(\ref{s3d1}) we find that
\begin{equation}
\label{real1}
\mathrm{Re}(S_n) = - \frac{1}{2}[s^{(0)}_n + s^{(1)}_n] + s^{(1)}_{nf}(0),    
\end{equation}
\begin{equation}
\label{imaginary1}
\mathrm{Im}(S_n) =  \frac{1}{2}[s^{(0)}_n + s^{(1)}_n] - s^{(1)}_{nf}(\pi/2).  
\end{equation}
This concludes our analysis of the QST applied to the two-qubit state encoded in the polarization and path DOF of a single photon.

\hfill

Now, some important remarks are in order. First, we call attention to the fact that creating polarization-path entangled states in a single photon is straightforward \cite{kok,englert}. For example, by passing a photon with diagonal polarization $\ket{D}$ through a PBS one generates the transformation $\ket{D} \rightarrow \frac{1}{\sqrt{2}}(\ket{H,0} + \ket{V,1})$, which results in a single-photon maximally entangled state. On the other hand, the implementation of entangling gates for two-qubits encoded in different photons is difficult because they do not interact directly, that is to say that it requires nonlinear couplings between photon paths \cite{klm}. Yet, we should note that to date entanglement involving different particles has found more significance in quantum technologies when compared to entanglement involving different DOF of a single particle \cite{erhard}. Second, QST implementations typically require reconfigurations of the measurement apparatus in order to account for the many observables to be measured. Here, our QST proposal is realizable only with a single experimental setup. Indeed, we observe that the SPM scheme described in Fig.~\ref{OPSPmeasurer} can be rearranged to work without the need to insert or remove the wave plates when measuring different observables. In doing so, one must add a 2:1 and a 1:1 beam splitter, together with two other PBS and two extra pairs of photodetectors, in the SPM apparatus (see Ref.~\cite{bayra}).

\hfill

It is also important to mention here that sources of error have to be considered in the experimental setup of Fig.~\ref{TPSPmeasurer}, e.g., the uncertainties in the angles of the wave plates used in the SPM. Errors in the experimental data provide a density matrix which may not correspond to a physical quantum state, i.e., the properties of unit trace or non negativity may not be fulfilled. For this reason, methods of statistical inference to fix the obtained unphysical states have been employed, such as maximum likelihood estimation (MLE) \cite{lvov}, and Bayesian inference \cite{lukens}. The application of the MLE method for the reconstruction of a two-qubit state, which is the present case, is explored in Ref.~(\cite{james}). This technique aims to find a physically plausible density matrix that maximizes the probability of obtaining the collected experimental data.

\section{Conclusion}

In conclusion, we have presented a QST method to read out the quantum state of two qubits encoded in the polarization and path DOF of a single photon. It consists in a single linear-optical setup in which all observables in the protocol can be assessed without the need of modifications in the apparatus. From the practical side, since the generation and control of polarization-path states are common tasks in quantum optical experiments \cite{kok2}, the tomographic reconstructions proposed here can represent an important step toward the application of such states in quantum technologies. From the theoretical side, we observed that the polarization-path density matrix can be completely characterized in terms of the OPSP related to each of the two possible paths taken by the photon, which are real numbers, and the TPSP, which are complex numbers that contain information about both the polarization and coherence of the photon. This result, to a certain extent, reveals the role played by the two-point Stokes parameters in quantum optics.     

As an outlook, given the increasing relevance of quantum optical experiments in simulating the behavior of open quantum systems \cite{liu,vid,somh}, and the simplicity with which information can be encoded on the polarization and path DOF of single photons, the QST protocol introduced here provides an acessible testbed for the study of the dynamics of open two-qubit systems \cite{viviescas,ghas,silt}. Indeed, practical experimental studies of decoherence or entanglement loss can be realized by preparing an ensemble of photons in a well-defined polarization-path state, passing them through the environment in question, and then reading out the output state with the setup described in Fig.~\ref{TPSPmeasurer}. In addition, with optical simulations of quantum thermodynamic processes, it is also possible to describe a qubit  system interacting with decohering and thermalizing environments by manipulation of the polarization and path DOF of the photons \cite{aguilar,khan}. Such simulations have significantly broadened our understanding about how coherence and system-environment quantum correlations affect the behavior of nonequilibrium quantum dynamics \cite{mancino,bellini,khan2}.  In this perspective, our QST technique for the reconstruction of single-photon polarization-path states can provide an extra tool for an even more detailed understanding of such processes.           
\section*{Acknowledgements}

The authors acknowledge support from Coordena{\c c}{\~a}o de Aperfei{\c c}oamento
de Pessoal de N{\'i}vel Superior (CAPES, Finance Code 001) and Conselho Nacional de Desenvolvimento Cient{\'i}fico e Tecnol{\'o}gico (CNPq). BLB acknowledges support from (CNPq, Grant No. 307876/2022-5 ).

\end{document}